\documentstyle[ppeuc,epsfig]{article}

\begin{document}

\title{Effect of a Collapsing Cluster on the CMB temperature and Power Spectrum}

\author{\addlink{Y. Dabrowski}{mailto:youri@mrao.cam.ac.uk}, M.P. Hobson,
A.N. Lasenby, C. Doran}

\institute{\addlink{Mullard Radio Astronomy Observatory}
{http://www.mrao.cam.ac.uk}, Cavendish Laboratory, Madingley Road,
Cambridge, CB3 OHE, UK}

\begin{abstract}
We present a new model for the formation of spherically symmetric
clusters in an expanding Universe. Both the Universe and the collapsing
cluster are governed by the same pressure less fluid equations for which
a uniform initial density profile is assumed. A simple perturbation
imposed on the initial velocity field gives rise to an over-density which closely
models real clusters. The computation of
photon paths allows us to evaluate the gravitational effects imprinted
on a Cosmic Microwave Background (CMB) photon passing through such an
evolving mass. We also consider the lensing properties of collapsing
clusters and investigate the effect of a population of such clusters
on the primordial microwave background power spectrum.
\end{abstract}

\section{Introduction}

There exist several mechanisms which produces anisotropies on the CMB:
the primary anisotropies which take place at the last scattering epoch
and the secondary anisotropies such as the Sunyaev-Zel'dovich effect
which occurs while the CMB photons are travelling through the universe.
We will concentrate here on a particular secondary anisotropy first
discussed by Rees \& Sciama (1968). As pointed out by these authors,
there is a gravitational effect on the CMB photons while they are
crossing evolving cosmic structures such as a collapsing cluster of
galaxies or expanding voids.
Indeed photons climb out of a slightly different potential well than the
one into which they entered. This effect is usually referred as the 
Rees-Sciama effect.\\

Early approaches to this problem were based on the ``Swiss Cheese''
model (\cite{rees-sciama}, \cite{dyer}, \cite{kaiser},
\cite{nottale82} \& \cite{nottale84}), whereas more recent attempts have
used the continuous Tolman-Bondi solution (\cite{panek}, \cite{saez}).
Our aim is to present here an improved model which, for the
first time, treats such a problem exactly (see section \ref{sec:model}).\\

In section \ref{sec:cluster} we apply our model to a rich galaxy cluster at a
redshift $z=0.08$. We investigate the properties of the collapsing
cluster and verify that its density-profile is of a realistic shape.
We then give, in section \ref{sec:DT}, an upper limit for
the temperature decrement that such an evolving cosmic structure
could imprint on the CMB. Our result is compared to previous works.\\

Section \ref{sec:dist} deals with distant clusters ($z=1$) since it is
reasonable to assume that, at such an epoch, those objects were in a stage
of formation and that our spherical collapse model would be suitable.
The focus is put on the gravitational lensing phenomenon and how this would
distort the observed CMB power spectrum.\\

We finish in section \ref{sec:discussion} by a short conclusion
and discuss the applicability of our work to model the microwave
decrement observed towards the, possibly lensed, quasar pair
PC1643+4631 A\&B.\\

The methods and results introduced in this poster are discussed in
further details in forthcoming papers (\cite{lasenby}, \cite{dabrowski}).

\section{Model}
\label{sec:model}
\begin{figure}[htbp]
\htmlimage{scale=1.5, thumbnail=0.5}
\centerline{\epsfig{file=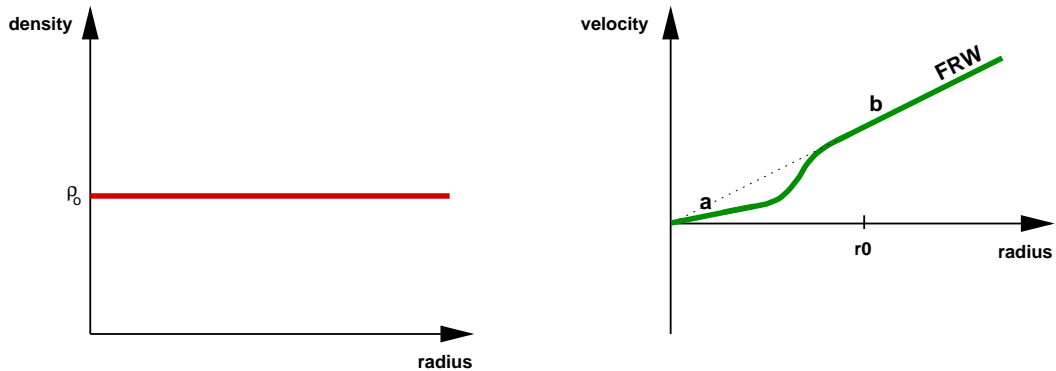,width=14cm}}
\caption[]{\label{fig:swiss2}Fluid initial conditions for our model.
Two parameters control the perturbation: $r_0$ and $a$.}
\end{figure}

In this poster we are concerned with the formation of spherical galaxy clusters
(i.e. pure infall) in an expanding background universe. As in the Tolman-Bondi
model applied by \cite{panek}, the evolution of both the collapsing cluster and
the expanding Universe are governed by the same pressure-less fluid equations.
Initially, the density of our fluid is uniform. Only the velocity field is perturbed
in order to form a realistic cluster. In a given cosmology, this perturbation is controlled
by two parameters ($r_0$ and $a$) as illustrated in figure \ref{fig:swiss2}. $r_0$ is related
to the size of the perturbation and $a$ to its rate of growth. Those parameters
are free and we choose them so that the resulting observed cluster is
realistic (see section \ref{sec:cluster}).\\

\begin{figure}[htbp]
\htmlimage{scale=1.5, thumbnail=0.5}
\centerline{\epsfig{file=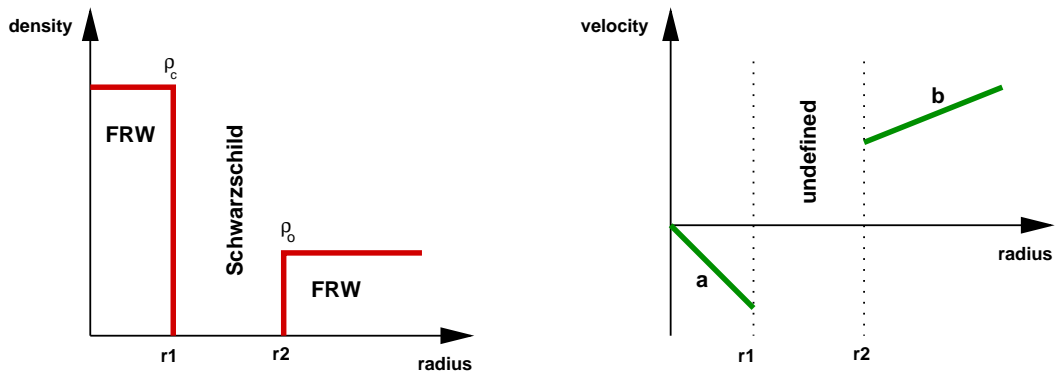,width=14cm}}
\caption[]{\label{fig:swiss1}Fluid initial conditions for Swiss-Cheese
models. Three parameters control the perturbation: $\rho_c$, $r_1$ and
$a$.}
\end{figure}

We find that our model has many advantages and treats the problem much more
rigorously than older attempts, particularly in the case of works based on the
``Swiss Cheese'' (SC) models. Indeed, as shown in figure \ref{fig:swiss1},
the SC models deals with discontinuous density and velocity distributions since an
unrealistic vacuum region is needed to separate the perturbation from the Universe.
Another improvement made with regard to the SC models is that our perturbation
is defined by only two free parameters rather than three in the SC case.

More recent studies of the problem use the Tolman-Bondi solution
(\cite{panek}, \cite{saez}), which gives a similar approach to our own. In particular,
both the forming cluster and the external Universe are treated as a whole and 
the fluid distributions are continuous. We believe our model is a
refinement of the work based on the Tolman Bondi solution. Our mathematical
formulation is clearer and does not need any approximation. The underling gauge
theory of gravity (\cite{grav}) that we use deals straightforwardly with observable
quantities such as the photon energy (\cite{lasenby}). Therefore, we do not need to
remove a posteriori dipole nor quadrupole-like anisotropy form the obtained
CMB fluctuation (e.g.  \cite{panek}).

\section{Cluster properties}
\label{sec:cluster}
In this section, we model the formation of a very rich Abell cluster. The characteristics
(see table \ref{table:cluster}) are chosen to be similar to those of the clusters
described in \cite{quilis} and \cite{panek} so that our results can be compared.\\

\begin{table}[htbp]
\htmlimage{scale=1.5, thumbnail=0.5}
\caption[Dummy]{\label{table:cluster}Cluster characteristics.}
\begin{center}
   \def\temp#1{\multicolumn{1}{|l||}{#1}}
   \begin{tabular}{l|c|c|}
   \cline{2-3}
   &$H_o=50$&$H_o=100$\\ \hline
   \temp{Redshift}&0.08&0.08\\ \hline
   \temp{Distance}&$450 {\rm \: Mpc}$&$225 {\rm \: Mpc}$\\ \hline
   \temp{Max density}&$0.7\times 10^4 {\rm \: p \, m^{-3}}$&$1\times 10^4 {\rm \: p \, m^{-3}}$\\ \hline
   \temp{Core radius}&$0.46 {\rm \: Mpc}$&$0.23 {\rm \: Mpc}$\\ \hline
   \temp{Mass $r<4 {\rm \: Mpc}$}&$1.3\times 10^{16} {\rm \: M_\odot}$&
             $5.8\times 10^{15} {\rm \: M_\odot}$\\
   \hline
   \end{tabular}
\end{center}
\end{table}

The cluster is placed at a redshift $z=0.08$, with a core radius $R_c=0.23 \, h_0^{-1} {\rm \: Mpc}$
($R_c$ is defined as the radius at which the cluster energy falls to one-half its maximum
value). The maximal baryonic density $\rho_{max}$ encounter by an observed photon
is taken to be $10^4\, h_0^{1/2} {\rm \: protons \, m^{-3}}$ and the baryon to dark-matter mass
ratio is assumed to be 0.1. The $h$-dependencies for $R_c$ and $\rho_{max}$ ensure that
the observed cluster characteristics (i.e. angular size and X-ray luminosity) are independent
of the Hubble parameter. Throughout this paper we assume $\Omega_0=1$.\\

\begin{figure}[htbp]
\htmlimage{scale=1.5, thumbnail=0.5}
\vspace{1cm}
\centerline{\epsfig{file=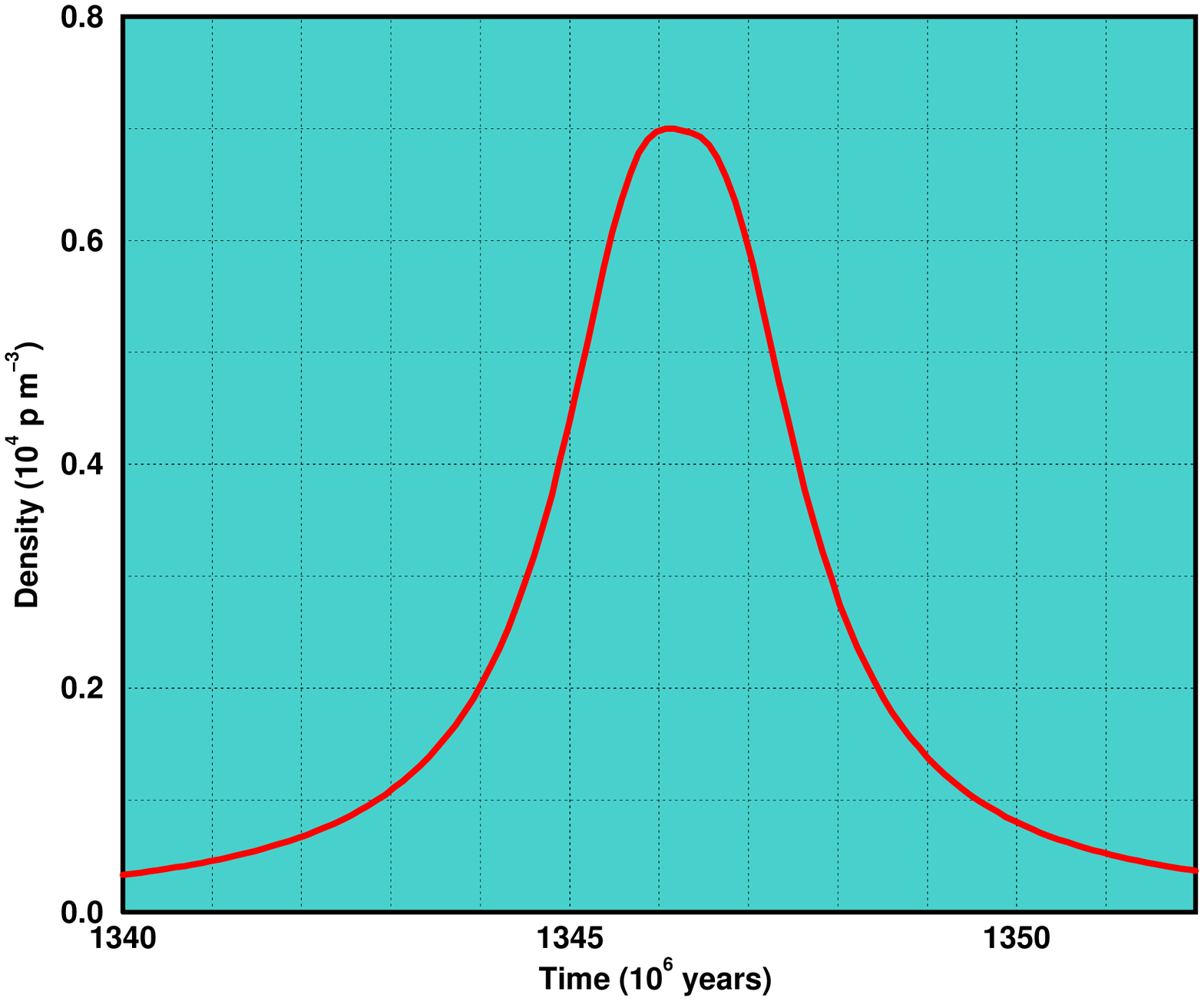,width=6.7cm}
                  \hspace{1cm}
	  \epsfig{file=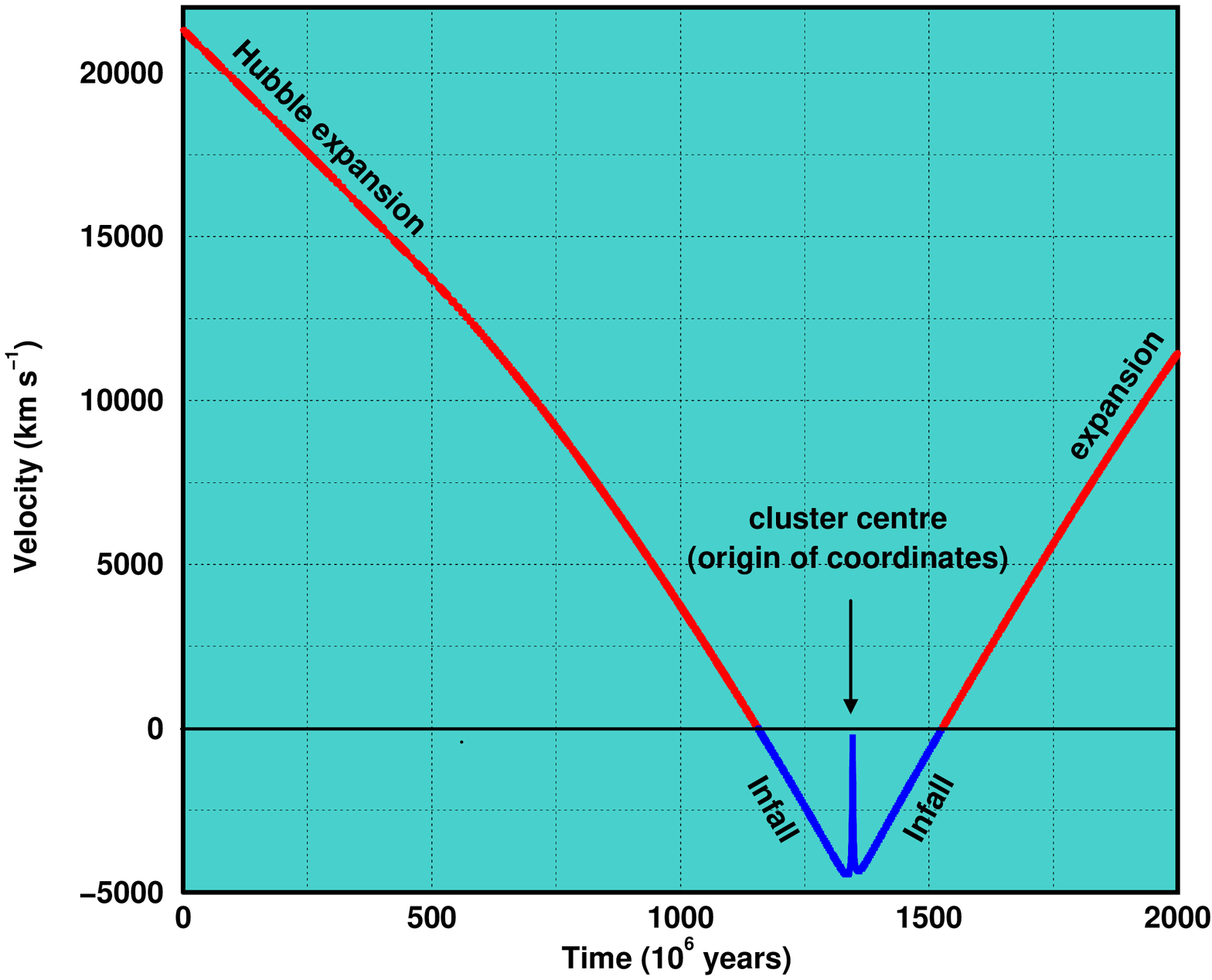,width=6.7cm}}
\caption[]{\label{fig:dens_vel} Plot of the fluid density (left) and fluid velocity (right) as
experienced by the photon while it is travelling through the Universe and the collapsing
cluster. The horizontal axis marks the time prior to the present epoch in millions of years.
We assume $h_0=0.5$.}
\end{figure}

Figure \ref{fig:dens_vel} shows the fluid density and the fluid velocity as a function of proper 
time as experienced by a photon which travels straight through the centre of the cluster.
From the figure we see clearly that the fluid distributions are continuous.

\begin{figure}[htbp]
\htmlimage{scale=1.5, thumbnail=0.5}
\vspace{1cm}
\centerline{\epsfig{file=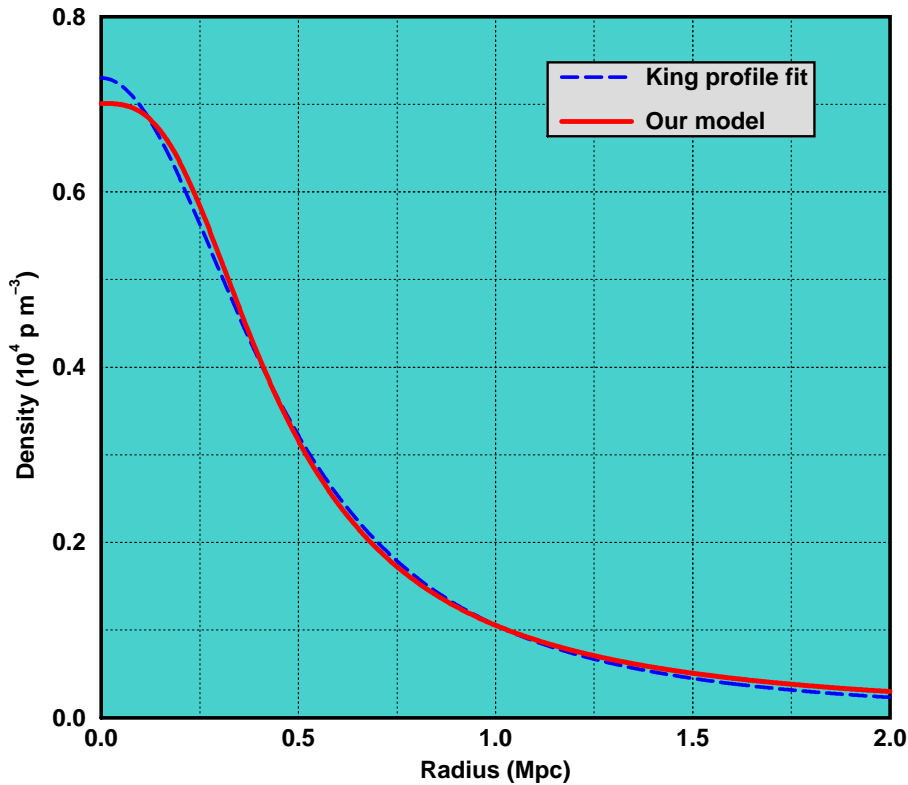,width=12cm}}
\caption[]{\label{fig:density} Plot of the density when the photon is
at the centre of the cluster. The obtained profile is realistic since
it can be fitted by a spherical King model.}
\end{figure}

We next consider (fig \ref{fig:density}) the density profile of the cluster at the
time $t_c=1436.1 {\rm \: Myr}$ ago, when the photon experienced the maximum
density $0.7\times 10^4 {\rm \: protons \, m^{-3}}$. In order to check that the obtained
density distribution is of realistic shape, we fit our profile to the equilibrium
spherical King model (\cite{king}):
\begin{equation}
\rho(r) = \rho_{\rm max}\left[1+\left(\frac{r}{R}\right)^2\right]^{-3\beta/2},
\label{king}
\end{equation}
where $\rho_{max}$ is the central density of the cluster, $R$ represents some core
radius and $\beta$ is the conventional power-law index for such models.
The results of a simple least-squares fit are
$\rho_{max}=0.73\pm 0.05 \times 10^4{\rm \: protons \, m^{-3}}$,
$R=0.52\pm 0.05 {\rm \: Mpc}$ and $\beta=0.8\pm 0.15$. The result of this fit
 is very encouraging since it shows that our model leads to a radial density
profile that matches quite closely those of observed clusters. 

\section{Effect on the CMB temperature}
\label{sec:DT}

We now consider photon energy and concentrate on the CMB anisotropy
produced when the photon passes through the cluster described in section \ref{sec:cluster}.
One must remember however that our model assumes a pressure-less fluid and so the cluster,
in pure-infall, might evolves too rapidly. Therefore the temperature perturbations given in
this section should be considered as upper limits.\\

\begin{figure}[htbp]
\htmlimage{scale=1.5, thumbnail=0.5}
\centerline{\epsfig{file=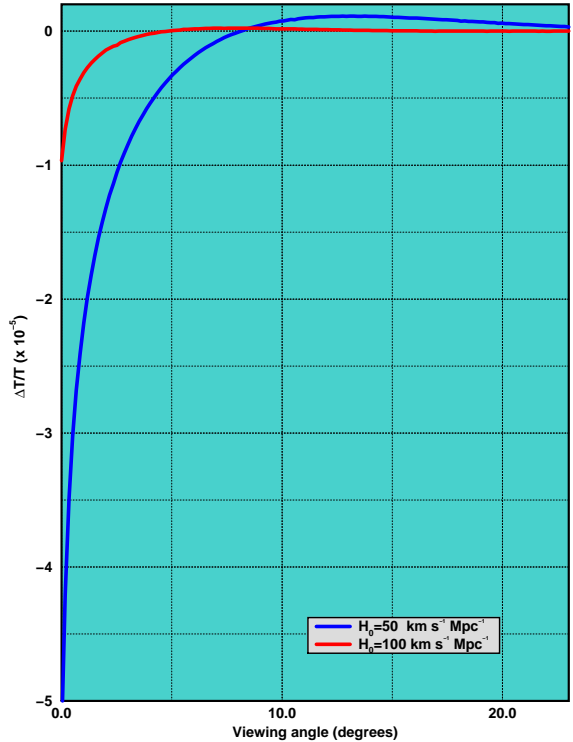,width=7cm}}
\caption[]{\label{fig:dt} Temperature perturbation $\Delta T/T$ imprinted on the CMB
as a function of the observed angle from the centre of the cluster.}
\end{figure}

Figure \ref{fig:dt} shows the CMB anisotropy due to the gravitational perturbation
of the cluster described in table \ref{table:cluster}. The maximum temperature
distortion occurs at the centre of the cluster and has the value $\Delta T/T=-0.96\times
10^{-5}$ and $\Delta T/T=-5.2\times 10^{-5}$ for $h_0=1$ and $h_0=0.5$ respectively.
One notices that the distortion extends to rather large projected angles (e.g.
$\Delta T/T \sim -1\times 10^{-6}$ for an observed angle of $\sim 2.5 {\rm \: degrees}$
in the $h_0=0.5$ case). We may also point out the fact that the anisotropy becomes
slightly positive at large angles.

We can compare the central decrements calculated above with those of previous
authors. For \cite{panek} type I and type II clusters, the calculated central decrements are
$\Delta T/T=-1.5\times 10^{-6}$ and $\Delta T/T=-6.0\times 10^{-6}$ respectively,
whereas \cite{quilis} quote the decrement $\Delta T/T=-1.2\times 10^{-5}$. For two
cluster models with similar physical properties \cite{chodorowski} finds
$\Delta T/T=-7.7\times 10^{-6}$. Finally \cite{nottale84} used the SC model to predict
a considerably larger central decrement of $\Delta T/T \sim 10^{-4}$. However, this
last value corresponds to a very dense, unrealistic cluster.
We can notice that, for $h_0=1$, our predicted value is in rough agreement with \cite{panek}
type II, \cite{chodorowski} and \cite{quilis}. However our result for $h_0=0.5$ is about
five times larger since previous works predict the same value whatever is $h_0$. We then
suggest that, in such a Universe, the effect on CMB photons may be more significant
that previously stated.

\section{Distant clusters}
\label{sec:dist}
We consider, in this section, the formation of a massive cluster located at a
redshift $z=1$. It is more reasonable to apply our model for a distant cluster
rather than the one described in section \ref{sec:cluster}. Indeed, at such
redshifts one could expect galaxy clusters to be in a formation process
and display large inward radial velocities. As from now, the cosmology
is taken to be $\Omega_0=1$ and $H_0=50 {\rm \: km \, s^{-1} \, Mpc^{-1}}$.
The cluster characteristics are the following: $z=1$,
${\rm maximum \: density} = 1.0\times 10^4 {\rm \:p \, m^{-3}}$,
${\rm core \: radius} = 0.46 {\rm \: Mpc}$ and the mass within $4 {\rm \: Mpc}$ is
$1.9 \times 10^{16} {\rm \: M_{\odot}}$.
In the next section we investigate the lensing effect caused by such a cluster
as well as the effect imprinted on the CMB temperature and power spectrum.

\subsection{Gravitational lensing and effect on the CMB}
\label{subsec:lensing}
\begin{figure}[htbp]
\htmlimage{scale=1.5, thumbnail=0.5}
\vspace{1cm}
\centerline{\epsfig{file=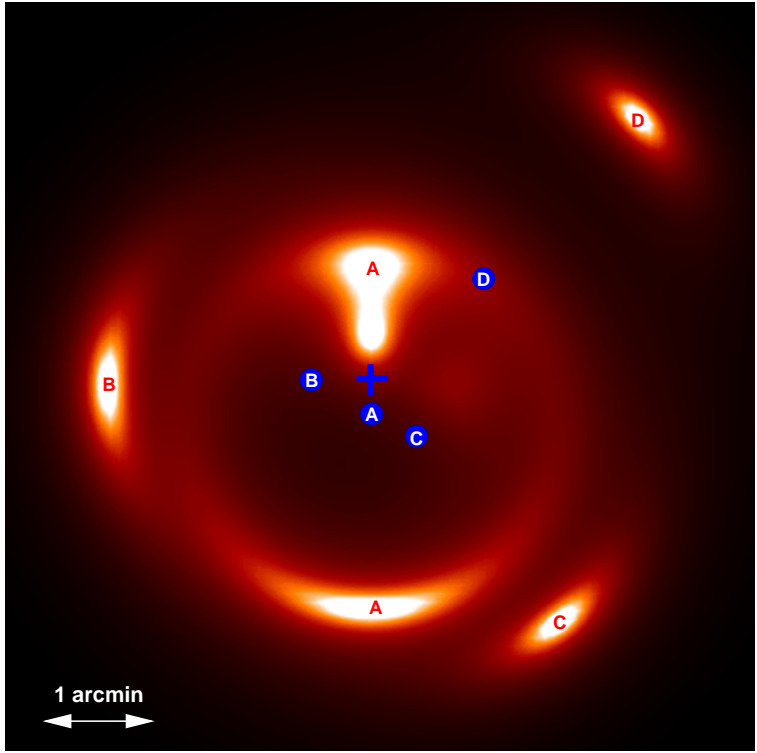,width=10cm}}
\caption[]{\label{fig:4lens}Lensing image of 4 sources at a redshift $z=3.8$ by the rich
cluster described in section \ref{sec:dist}. The blue bullets (A,B,C,D) show the positions
of the sources as seen in the absence of the cluster. The blue cross marks the centre of
the cluster. The source A placed on the lens caustic displays two images, one radially
and the other tangentially elongated.}
\end{figure}

\begin{figure}[htbp]
\htmlimage{scale=1.5, thumbnail=0.5}
\vspace{1cm}
\centerline{\epsfig{file=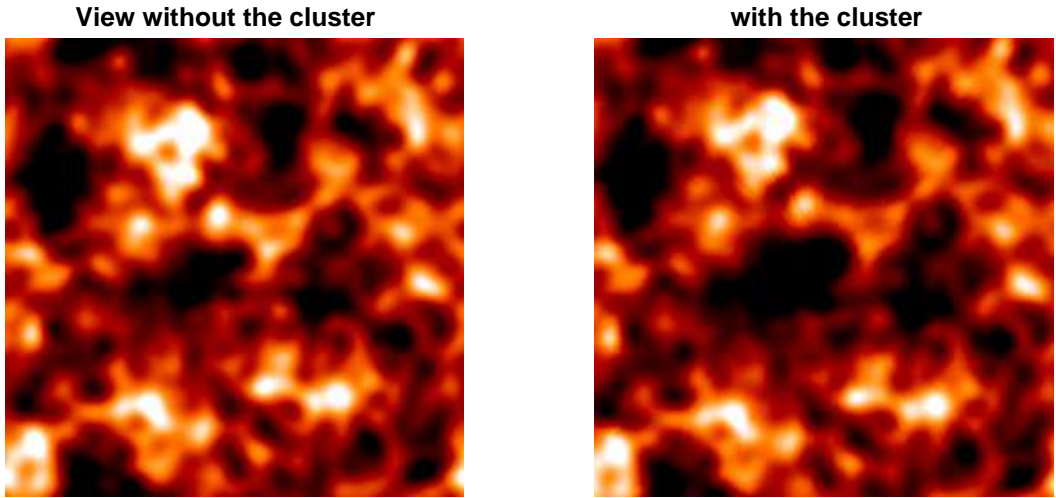,width=14cm}}
\caption[]{\label{fig:cmb} Effect of the collapsing cluster described in section \ref{sec:dist}
on the CMB fluctuations. The first map is a simulation of the CMB fluctuations due to
inflation. The second one represents the same patch of the sky (4x4 degrees) but with a
rich cluster of galaxies at its centre.}
\end{figure}

A massive cluster such as described above is a powerful gravitational
lens. This effect is illustrated in figure \ref{fig:4lens}.
Our model also allows a quantitative study of the effect of
dynamically evolving lenses on the CMB fluctuations. By simulating maps of
fluctuations in the CMB due to inflation and/or topological defects,
we can investigate how features in these maps are affected by the
presence of the lens. As an example, figure \ref{fig:cmb} shows the effect of our collapsing
cluster on a CMB fluctuations realisation. The result is a central decrease of
temperature at the centre (i.e. $180 \:  \mu{\rm K}$ in our case) as well as a global
magnification of the features.

\begin{figure}[htbp]
\htmlimage{scale=1.5, thumbnail=0.5}
\vspace{1cm}
\centerline{\epsfig{file=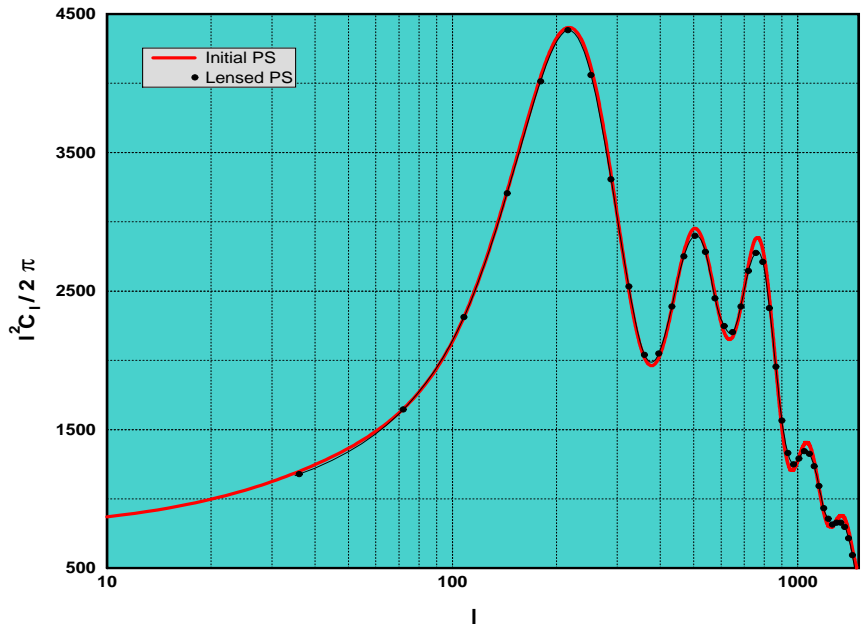,width=11cm}}
\caption[]{\label{fig:power}Effect on the CMB power spectrum of the rich cluster described
in section \ref{sec:dist}. We assume one cluster per 10x10 degrees field. The red curve is the
unperturbed power spectrum as the black dots represent the ``lensed'' spectrum.}
\end{figure}

Aside from studying individual maps, we can also
study the statistical effect that a population of such lenses would
have on the power spectrum of the CMB fluctuations. This power
spectrum is currently of immense theoretical and observational
interest, since it is now a possibility that the spectrum may be
measured, and so the values of cosmological constants may be 
found to unprecedented accuracy. Such determinations depend, however,
on distinctive features in the power spectrum which may be affected by a
population of large clusters. Previous work on this effect has been performed
such as  \cite{seljak}, but we have now incorporated the effect of dynamically
collapsing clusters in a proper general relativistic fashion. Figure \ref{fig:power}
gives the result obtained with the cluster used in the current section.

We find that the distinctive peaks in the power spectrum are slightly smoothed out by
such a population, and so this effect should be taken into account in
future analyses.

\section{Discussion}
\label{sec:discussion}

A CMB decrement of $\sim 380 \: \mu {\rm Jy}$ has been observed by the Ryle-Telescope
towards the pair of quasars PC1643+4631 A \& B at red-shift $\sim 3.8$ separated by 100
arc-seconds (\cite{jones}).
Assuming a Sunyaev-Zel'dovich effect from a cluster of galaxies, this is indicative of a
rich intervening cluster of total mass $\sim 10^{15} {\rm \: M_{\odot}}$, although,
no X-ray cluster has yet been observed in that direction. This suggests that the object
is lying at a red-shift greater than one.
The quasars' spectra analysis reveals that they might be originating from a unique
source lensed by the distant cluster (\cite{saunders}).

Using our model, we are able to fit those observations and retrieve the S-Z flux
decrement as well as the 100 arc-seconds separation. To do so, we have to consider the
cluster described in section \ref{sec:dist} with an electron temperature of $2\times10^7 {\rm \: K}$,
which corresponds to a rich but rather cool cluster.
As we stated in section \ref{subsec:lensing}, the resulting Rees-Sciama temperature
decrement is $\leq 180 \: \mu {\rm K}$ which is a considerable fraction of the  S-Z decrement,
here $\sim 450\: \mu {\rm K}$.
Fitting those data is of great importance since such a distant lensed-pair has previously
never been observed. As the cluster lies at a rather large red-shift, we believe that a model
such as ours, computing rigorously both the cluster and the universe evolutions, is
needed.\\

\end{document}